\documentclass[aps, twocolumn, superscriptaddress, showpacs, nofootinbib, longbibliography]{revtex4-1}

\usepackage[utf8]{inputenc}
\usepackage[T1]{fontenc}
\usepackage{ae,aecompl} 
\usepackage{graphicx}
\graphicspath{{figures/}} 
\usepackage{amsmath}
\usepackage{color}
\usepackage{amssymb}
\usepackage{latexsym}
\usepackage{wasysym}
\usepackage{psfrag}
\usepackage{ifthen}
\usepackage[citecolor=blue,colorlinks=true]{hyperref}
\usepackage{longtable}
\usepackage{float}
\usepackage[utf8]{inputenc}
\usepackage{lineno}
\usepackage{units}
\usepackage{multirow}
\usepackage{gensymb}

\newcommand{\answer}[1]{{\color{black}#1}}

\newcommand{\msun}{$M_{\odot}$}

\newcommand{\apjl}{Astrophys. J. Letter} 

\begin{document}
\title{A long-duration transient, gravitational-wave search pipeline}

\author{A. Macquet}
\affiliation{Artemis, Universit\'e C\^ote d’Azur, Observatoire de la C\^ote d’Azur, CNRS, Nice 06300, France}

\author{M.A. Bizouard}
\affiliation{Artemis, Universit\'e C\^ote d’Azur, Observatoire de la C\^ote d’Azur, CNRS, Nice 06300, France}

\author{N. Christensen}
\affiliation{Artemis, Universit\'e C\^ote d’Azur, Observatoire de la C\^ote d’Azur, CNRS, Nice 06300, France}

\author{M. Coughlin}
\affiliation{School of Physics and Astronomy, University of Minnesota, Minneapolis, Minnesota 55455, USA}

\begin{abstract}

As the sensitivity and observing time of gravitational-wave detectors increase, a more diverse range of signals is expected to be observed from a variety of sources. Especially, long-lived gravitational-wave transients have received interest in the last decade.
Because most of long-duration signals are poorly modeled, detection must rely on generic search algorithms, which make few or no assumption on the nature of the signal. However, the computational cost of those searches remains a limiting factor, which leads to sub-optimal sensitivity. Several detection algorithms have been developed to cope with this issue.
In this paper, we present a new data analysis pipeline to search for un-modeled long-lived transient gravitational-wave signals with duration between $10-10^3$ s, based on an excess cross-power statistic in a network of detectors. The pipeline implements several new features that are intended to reduce computational cost and increase detection sensitivity for a wide range of signal morphologies. The method is generalized to a network of an arbitrary number of detectors and aims to provide a stable interface for further improvements.
Comparisons with a previous implementation of a similar method on simulated and real gravitational-wave data show an overall increase in detection efficiency for all but one signal morphologies tested, and a computing time reduced by at least a factor $10$.
\end{abstract}

\maketitle

\section{Introduction}

A new era in astronomy began in September 2015 with the observation of gravitational waves (GWs) from the merger of two stellar mass black holes~\cite{Abbott:2016blz}. Since then, the Advanced LIGO~\cite{Aasi:2014jea} and Advanced Virgo~\cite{Acernese:2014hva} have regularly observed a larger volume of the Universe leading, among major discoveries, to the observation of the merger of two neutron stars~\cite{Abbott:2017qsa} in August 2017 associated to gamma ray burst GRB190817A~\cite{Abbott:2017mdv} followed up kilonova AT2017gfo in NGC4993~\cite{Coulter:2017wya}.
As of mid of 2021, Advanced LIGO and Advanced Virgo have reported $\sim$ 50 confirmed mergers of compact objects, black holes and/or neutron stars~\cite{Abbott:2020niy}.

Yet many GW sources have not yet been observed: core collapse supernova, isolated neutron stars, magnetars, cosmic strings, and the resulting stochastic background of GWs~\cite{Bailes:2021tot}. The diversity of the GW signal expected from these sources require different detection algorithms. When the GW signal waveform is predicted analytically, matched filter techniques can be used. In practice, this concerns mainly compact objects binary coalescence, cosmic strings signals~\cite{Damour:2001bk} and GWs from pulsars \cite{Sieniawska:2019hmd,Glampedakis2018}.
When the GW emission is poorly modeled, detection will rely on unconstrained searches that make few assumptions about the characteristics of the signal. In the last twenty years, several search algorithms have been developed, mostly focusing on GW signals of duration less than a few seconds ~\cite{Anderson:2000yy,Klimenko:2015ypf, Chatterji:2004qg, Sutton:2009gi, Cornish:2015}. More recently, transient GW signals of longer duration have received attention, bridging the gap between short-duration transient and continuous emission of GWs, and dedicated search algorithms have been developed \cite{Prix:2011qv, Thrane:2010ri, Thrane:2015psa, Thrane:2015wla, Keitel:2018pxz, Miller:2018rbg, Oliver:2019ksl}. 

Several astrophysical processes could be at the origin of long-duration transient GWs emission, for example, those related to core collapse supernova, compact object binary mergers and isolated neutron stars ~\cite{Thrane:2010ri}. Some of them are associated to the most energetic phenomena observed in the Universe.
There is evidence~\cite{Woosley:2006fn,Nagataki:2018} that core collapse supernovae and long gamma-ray bursts (GRB) are connected to the death of massive stars where the iron core collapses under its own gravity, forming either a black hole or a highly magnetized neutron star, releasing an incredible amount of energy (\unit[$10^{53}$]{erg}) mainly through neutrino emission, while $\sim$ 1\% goes into the kinetic energy of the explosion~\cite{Janka:2012wk}. Once the collapse is triggered, very powerful non-spherical flows develop in the outer region of the proto-neutron star that are expected to generate GWs energetically bounded to \unit[$10^{44}-10^{47}$]{ergs}~\cite{Abdikamalov:2020jzn}. The GW emission will last until the onset of the explosion or until a black hole is formed. The signal is expected to be no longer than \unit[$1-2$]{s}.



In the collapsar model, massive stars collapse to black holes either without an initial supernova explosion or via fallback accretion after a successful but weak explosion~\cite{MacFadyen:2001}; a rotating black hole is formed while the inner layers of the star lacks momentum to eject all the matter. Over a period of minutes to hours, \unit[$0.1-5$]{\msun} falls back onto the collapsed remnant, turning it into a black hole and establishing an accretion disk. GWs may be emitted by disk turbulence and disk instabilities that may lead to clumping or disk fragmentation~\cite{Piro:2006ja, vanPutten:2001sw}. The GW signal expected from accretion disk fragmentation would last $\unit[\mathcal{O}(10-100)]{s}$ with a characteristic strain $h \sim 10^{-22}$ at $\unit[100]{Hz}$ for a source at \unit[100]{Mpc} \cite{Piro:2006ja}. 
When the core collapse explosion is successful, a magnetar is formed. Convective currents and dynamical and secular nonaxisymmetric rotational instabilities in the proto-magnetar develop and may emit GWs~\cite{Bucciantini:2009vp}. In both scenarios, a GRB jet is launched either thanks to magnetohydrodynamical processes and neutrino pair annihilation powered by accretion or by the high Lorentz factor outflow that follows the birth of the proto-magnetar.

When a magnetar is formed, gravitational wave emission from viscosity-driven "spin-flip"  instability may last hours to days, with a detection horizon of $3-4$ Mpc for Advanced LIGO/Advanced Virgo detectors and unmodelled searches \cite{2018MNRAS.480.1353D,Dallosso:2021wzr, 2018PhRvD..98d3011S}.

The merger of two neutron stars will form a hot supermassive neutron star; depending on the component masses, the centrifugal forces induced by differential rotation and the stiffness of the nuclear equation of state may allow it survive for hundreds of milliseconds before collapsing to a black hole or form a massive neutron star~\cite{RaLa2014,Sarin:2020gxb}. It is very likely that the remnant is surrounded by an accretion disk that may endeavor instabilities like in the collapsar scenario. 
If the newly formed neutron star survives more than a few seconds, it could emit long-lived GW through magnetic field-induced ellipticity ~\cite{BoGo1996, Cut2002} or $r$-mode instabilities~\cite{LiOw1998}, although the precise amplitude of such signals remains unclear. So far, no post-merger GW signals have been detected for any of the binary neutron star mergers found in LIGO and Virgo data~\cite{Abbott:2017dke,Abbott:2020aai, AbEA2018b}. 

Isolated neutron stars are another potential source of long-duration GW signals. Sudden speed-ups of the rotation of pulsars observed in radio and X-ray data are followed by a period of relaxation (weeks long) during which the pulsar slows down. GWs may be emitted during this period but the amplitude is expected to be low as the rotational energy changes remain below \unit[$10^{43}$]{erg}~\cite{Krawczyk:2003my, van_Eysden_2008, Bennett_2010, Melatos_2015}.
Seismic phenomena in the crust of magnetars are thought to be at the origin of soft gamma repeaters and anomalous X-ray pulsars. Soft gamma repeaters giant flares are associated to huge emission of electromagnetic energy, up to \unit[$10^{46}$]{erg}, followed by long duration quasi periodic oscillations which may be associated to GW emission over the same time scale ~\cite{Ioka:2000hs,2011PhRvD..83j4014C,Quitzow_James_2017}. The recent observation of GRB 200415a, suggesting that magnetar giant flare may be a distinct class of short GRB, with a substantially higher volumetric rate than compact object mergers \citep{2021ApJ...907L..28B}, is re-enforcing the interest for magnetar giant flare events in nearby galaxies.

The diversity of long transient expected GW waveforms has lead to the development of algorithms that do not rely on a signal model. Coherent waveburst~\cite{Klimenko:2015ypf, drago2021coherent} and X-pipeline~\cite{Sutton:2009gi}, used for short-duration searches, have been adapted to search for transients with duration up to a few hundred of seconds, while the \texttt{STAMP} excess cross-power algorithm~\cite{Thrane:2010ri} has been developed to target specifically long and very long transient signals lasting up to several weeks. It has been used to search for long duration GW transients associated to GRBs~\cite{Aasi:2013cya}, for post-merger GW signals associated to GW170817~\cite{Abbott:2017dke, AbEA2018b} and adapted to perform an all-sky/all-time search for long duration GW transient in LIGO and Virgo data~\cite{Abbott:2015vir, Abbott:2017muc, Abbott:2019heg}.

An enhanced version of the \texttt{STAMP} algorithm is presented in this article. It is a complete rewrite in python of the all-sky/all-time \texttt{STAMP-AS} pipeline that was built using the \texttt{STAMP} algorithm library written in Matlab. As such, it has been optimized to search for GW signals \answer{of duration in the range $\unit[10-10^3]{s}$} in a large data set at a lesser computing cost than \texttt{STAMP}. It especially implements a hierarchical strategy, similar to the algorithm proposed in~\cite{Thrane:2015psa} to select the most interesting periods of the data without loosing detection efficiency.

This paper is organized as follows. In section~\ref{sec:methods}, we present the formalism of the analysis and the methods used to generate candidate events in the framework of a 2 detector search. We describe the implementation of the pipeline and the methods used for background and efficiency estimation in Section~\ref{sec:pipeline}. Section~\ref{sec:performances} summarizes the performance of the pipeline over simulated Gaussian noise and real data from the second LIGO-Virgo observation campaign (O2). Finally, we summarize those results in section~\ref{sec:conclusion} and propose several improvements to increase the pipeline sensitivity in the future.

\section{Overview of a cross-correlation GW transient search algorithm}
\label{sec:methods}

\subsection{Definitions and conventions}

We are considering a network of GW detectors whose strain data time-series $s(t) = n(t) + h(t)$ is a linear sum of independent detector noise $n(t)$ and the detector's response to a GW strain amplitude given by $h(t)$. The detector noise is itself the sum of random noise and non-Gaussian noise transients, or "glitches". The GW signal is assumed to be described by two polarization modes, $h_+(t)$ and $h_{\times}(t)$ and originates from a point-like source whose sky-position is given 
by the right ascension and declination ($\alpha, \delta$). We define $\hat{\Omega}$ as the direction to the source and $\tilde{h}(f)$ the Fourier transform of any $h(t)$ time-series. The detector's response to a GW strain is the linear combination of the two polarisations weighted by the detector antenna factors $h(t) = F^+(t, \hat{\Omega}) \times h_+(t) + F^{\times}(t, \hat{\Omega}) \times h_{\times}(t)$. We consider an interval of duration $T$ of GW strain data that are discrete measurements sampled at $f_s$. In the following, the variable $t;$ refers to the time segment start time.\\

The \texttt{STAMP} algorithm is an extension of the radiometer method developed to detect point-like sources of stochastic background GWs~\cite{Ballmer:2005uw}. To estimate the GW strain power spectrum of a transient signal, excess power is searched in frequency-time maps ($ft$-maps) formed by cross-correlating the data of two spatially separated gravitational wave detectors $I$ and $J$. Following~\cite{Thrane:2010ri} an estimator of the GW power in a single $ft$-pixel is given by

\begin{equation}
\hat{Y}(t;f,\hat{\Omega}) \equiv \textrm Re[ Q_{IJ}(t;f, \hat{\Omega}) \,\tilde{s}_I^{\star}(t;f)\,\tilde{s}_J(t;f)]
\end{equation}
where
\begin{equation}
Q_{IJ}(t;f,\hat{\Omega}) = \frac{1}{\epsilon_{IJ}(t; \hat{\Omega})} e^{2 \pi i f \hat{\Omega} \cdot \Delta \vec{x}_{IJ}/c}
\end{equation}
is a filter function that takes into account the arrival time delay of the signal in the two detectors
whose distance is given by $\Delta \vec{x}_{IJ}$ and the pair efficiency
\begin{equation}
\epsilon_{IJ}(t; \hat{\Omega}) \equiv \frac{1}{2}\left(F_I^{+}(t;\hat{\Omega}) F_J^+(t;\hat{\Omega}) + F_I^{\times}(t;\hat{\Omega})F_J^{\times}(t;\hat{\Omega})\right)
\end{equation}
which weights the GW strain cross-power according to the alignment of the detectors.
To normalize the cross-correlation, we compute the variance of $\hat{Y}$ for which an estimator is 
\begin{equation}
\hat{\sigma}_{Y}^2 (t;f,\hat{\Omega}) = |Q_{IJ}(t;f,\hat{\Omega})|^2 P_I(t;f)P_J(t;f)
\end{equation}
where $P_I(t;f)$ is the noise one-sided auto-power spectrum. We then define the signal-to-noise ratio $\textrm{SNR}(t;f,\hat{\Omega})$ for a single pixel
\answer{
\begin{equation}
\label{eq:SNR}
\begin{split}
  \textrm{SNR}(t;f,\hat{\Omega})
  & \equiv  \frac{\hat{Y}(t;f,\hat{\Omega})}{\hat{\sigma}_Y (t;f,\hat{\Omega})} \\
  & = \textrm{Re}\left[\frac{\tilde{s}_{I}^{\star}(t;f)\tilde{s}_{J}(t;f)} { \sqrt{P_I(t;f)P_J(t;f)}} ~ e^{2 \pi i f \hat{\Omega} \cdot \Delta \vec{x}_{IJ}/c} \right].
    \end{split}
  \end{equation}
 }
$\textrm{SNR}(t;f,\hat{\Omega})$ depends only on the single-detector whitened statistic
\begin{equation}
  \tilde{y}_I(t;f) \equiv \frac{\tilde{s}_I(t; f)}{\sqrt{P_{I}(t;f)}}
\end{equation}
and the time delay $\tau \equiv \hat{\Omega} \cdot \Delta \vec{x}_{IJ}/c$ of the signal in the two detectors.

In the context of an all-sky search, the source direction $\hat{\Omega}$, and therefore $\tau$, are unknown. An error in the time delay induces a dephasing in the computation of $\hat{Y}(t;f,\hat{\Omega})$ that can cause an underestimation of the SNR of coherent signals. 
A solution is to span all sky-positions $\hat{\Omega}$ and retain the one that gives the largest SNR. 
That was the strategy implemented in \texttt{STAMP-AS} used to search for long duration transient GW signals in initial LIGO
data~\cite{Prestegard:2016, Abbott:2015vir} and advanced LIGO data~\cite{Abbott:2017muc, Abbott:2019heg}. However, the computational time required to process numerous sky positions was a limitation of the pipeline. Besides, background estimation requires repeating, a large number of times, the same coherent cross-correlation of the data streams for each sky position tested using complete $ft$-maps while a large fraction of the pixels do not contain relevant information. As a consequence the amount of simulated background was restricted to $\sim 100$ years, and the number of sky positions tested was limited to a few. All these sub-optimal features resulted in a loss of sensitivity of $\sim 10-20 \%$ ~\cite{Abbott:2015vir}.

The \texttt{PySTAMPAS} pipeline addresses these limitations
by implementing the hierarchical approach proposed in~\cite{Thrane:2015psa} which consists of first identifying the most interesting clusters of pixels in single-detector auto-power $ft$-maps. In a second stage, a coherent detection statistic is computed using only the pixels that have been selected in the first stage. The computationally intensive calculations are carried out only once, allowing rapid background estimation without sacrificing the sensitivity gained by the use of coherence and spanning the whole sky positions.
The gain in computational performance has also allowed the introduction of the use of several time-frequency resolutions to gain sensitivity to GW signals that may have time-varying frequency evolution. In the following sections, we describe the different computations that are performed at each stage.

\subsection{Single detector stage}
\label{sec:single}
\subsubsection{Single detector \textit{ft}-map} 

The simplest time-frequency representation of the GW time series $s_I(t)$ is a spectrogram obtained using one-sided Fourier transform of short segments of duration $\Delta t$. The short segments are first Hann-windowed and overlap by 50\% with each other such that the pixels resolution is respectively $\left(\Delta t /2\right) \times \left(1/\Delta t\right)$  in time and frequency -- the factor $\frac{1}{2}$ comes from the $50\%$ overlap between short segments.
  
The spectrograms are whitened by the one-sided power spectral density of each segment $P_{I}(t;f)$. Two methods to compute the auto-power have been implemented. The first one, inherited from \texttt{STAMP}, takes the average of $|\tilde{s}_{I}(t;f)|^2$ over time-neighboring pixels in a similar way to the Welch's method. The other one considers the median over the frequency-neighboring pixels. The pros and cons of the two methods are discussed in section \ref{subsec:PSD}. For each time-frequency resolution, $ft$-maps of the whitened statistic $\tilde{y}_I(t;f) $ are built.

The duration $\Delta t$ of the \answer{Fourier transformed segments} is an arbitrary choice that depends of the type of signal searched. Long-duration GW searches generally use \answer{Fourier transformed segments} of duration $ \simeq 1 \rm s$ which are suited to reconstruct signals lasting $\sim 10^1 - 10^3 \rm s$. However when the frequency evolution of the signal is changing with time, parts of it can be better reconstructed using different resolutions. In order to improve signal reconstruction as demonstrated by coherent waveburst~\cite{Klimenko:2015ypf}, we opt for a multi-resolution approach which consists in building several $ft$-maps of different resolutions and combine them into a single, multi-resolution $ft$-map.
 
\subsubsection{Clustering}
The long-duration GW signal signature in $ft$-maps \answer{appears as a cluster of pixels} that a pattern recognition algorithm must be able to \answer{reconstruct} without assuming a model. A year-long data set is typically used. 
The unknown morphology assumption leads us to consider a seed-based clustering algorithm. \answer{The principle is to group high-energy pixels together by proximity, without imposing any preferred morphology for the cluster}. For \texttt{PySTAMPAS}, we have adapted the \texttt{burstegard} algorithm, developed for \texttt{STAMP}~\cite{Prestegard:2016} to multi-resolution $ft$-maps.

We consider all \answer{pixels $\tilde{y}_{I}(t;f)$ from every map with individual resolution $\Delta t_i \times \Delta f_i$.} Pixels for which $|\tilde{y}_{I}(t;f)|$ exceeds a given threshold are kept to form a set of pixels for which we keep the time and frequency of the bottom left corner, $\Delta t_i$, $\Delta f_i$, and $\tilde{y}_{I}(t;f)$. The clustering algorithm starts with a seed pixel, the first pixel in the list, as the order does not matter. All pixels that are \answer{above threshold and} within a given distance \answer{(in time and frequency)} of the seed become part of the same cluster, \answer{whatever their resolution}. Each new pixel added to the cluster is then becoming the seed pixel and the same process is repeated recursively until no more pixels can be added. The next remaining unclustered pixel becomes the seed of the next cluster and the operations are applied again until all isolated pixels have been clustered.
To eliminate clusters composed of only a few pixels, we select clusters that have a user-determined minimal number of pixels. The different parameters of the clustering (pixel threshold, radius and minimal number of pixels per cluster) are free parameters that can be tuned considering that the number of operations is proportional to $\mathcal{O}(N \log(N))$), where $N$ is the number of pixels above threshold. As the GW signal energy is spread over many pixels, the threshold on $|\tilde{y}_I(t;f)|$ should not be too selective, and the distance between 2 pixels should not be too strict as well. 



\subsection{Coherent analysis}
\label{sec:coherent}
Considering all possible detector pairs, clusters from one detector are cross-correlated with the other detector's pixels. At this stage, the clusters can be composed of pixels of different time-frequency resolution. To be able to cross-correlate pixels of different time-frequency resolution, we define virtual pixels that have resolution $\min{\Delta t_i} \times \min{\Delta f_i}$. Each of these pixels is assigned a value that is the largest $|\tilde{y}_I(t;f)|$ value of all pixels that overlap the virtual pixel. As $\Delta t_i \times \Delta f_i$ is constant over all resolution maps, the virtual pixels assigned values have the same weight. The same construction of virtual pixels is performed for the pixels of the other detector's $ft$-map. 

The cross-correlation SNR given by Eq.~(\ref{eq:SNR}) is then computed considering the virtual pixels. As already mentioned, pixel SNR depends only on the time delay between two detectors $\tau = \hat{\Omega} \cdot \Delta\vec{x}_{IJ} /c$. In an all-sky search, the direction to the source is not known \textit{a priori}, and an error on the time delay can cause to underestimate the SNR of coherent signals. A solution is to span the time delay parameter space over all possible values for a given pair. The maximal SNR loss due to an error of $d\vec{\Omega}$ corresponding to $d\tau$ is
\begin{equation}
\textrm{SNR}(t;f,\hat{\Omega} + d \vec{\Omega}) = \cos(2\pi f d\tau)\, \textrm{SNR} (t;f,\hat{\Omega})
\end{equation}
The time delay bin size $d\tau$ is determined such that the maximal SNR loss is lower than $\epsilon \in [0~ 1]$ for the maximal frequency considered in \answer{each cluster}. 
\begin{equation}
\label{eq:dtau}
  d\tau = \frac{\arccos(\epsilon)}{2 \pi f_{max}}
\end{equation}
with $f_{max}$ being the maximal frequency of all pixels of the cluster, which can be much lower than the maximal frequency of the search, reducing the number of time delays to test.
In the most general case, we would need to test $N_{\tau}$ time delay values between 0 and $\Delta x_{IJ} /c$ by steps of $d\tau$ to \answer{recover} a signal accurately. However, because we are considering a phase factor, a degeneracy appears:
$\epsilon = \cos(2 \pi f d\tau) = \cos(2 \pi f (d \tau + 1/f))$. As a consequence, for a pixel at frequency $f$, it is sufficient to test time delays in the interval $[0, 1/f]$ instead of $[0, \Delta x_{IJ} /c]$ to get the correct phase factor. In the case of a broadband cluster with pixel frequencies between $f_{min}$ and $f_{max}$, this interval is the largest for $f = f_{min}$, so we need to test time delay values between 0 and $1 / f_{min}$ by steps of $d\tau$. Finally, the number of time delays to test to \answer{recover} a signal with an accuracy $\epsilon$ is 
\begin{equation}
\label{eq:ntau}
N_{\tau} = \frac{2 \pi}{\arccos(\epsilon)}\frac{f_{max}}{f_{min}} ~.
\end{equation}
In the end, $N_{\tau}$ remains rather small (less than a few hundred), especially compared to the thousands of sky positions that need to be tested using a regular grid in sky coordinates $\alpha$, $cos(\delta)$. Since the computations are done only over the small subset of pixels that constitute the cluster, it is possible to test hundreds of time delays in a reasonable time and therefore limit the loss of SNR to $\epsilon = 0.95$ regardless of the signal morphology as shown in section \ref{subsec:skylocation}.


The time delay $\tau_0$ that maximizes the sum of all pixel SNRs provides a detection statistic that reflects the significance of the cluster 
\begin{equation}
\label{eq:SNRgamma}
\textrm{SNR}_{\Gamma} \equiv \sum \limits_{(t,f) \in \Gamma} \textrm{SNR}(t;f,\tau_0).
\end{equation}
This detection statistic is used to test the hypothesis of a GW signal or the null hypothesis. However, hierarchical processing methods such as \texttt{PySTAMPAS} applied on real GW data tend to bias the selection of triggers because of the presence of noise outliers in one detector. When combined with noise fluctuation in the second detector, such triggers may have large $\textrm{SNR}_{\Gamma}$ values despite being incoherent. In order to mitigate this effect we estimate the residual noise energy that is left in one detector's data after subtracting the sum of $|\tilde{y}_I(t;f)|^2$ over all pixels belonging to the cluster. We define the quantity
\begin{equation}
  E^{res}_I \equiv \sum \limits_{(t,f) \in \Gamma}|\textrm{SNR}(t;f) - |\tilde{y}_I(t;f)|^2|.
\end{equation}
For a coherent GW signal, \answer{recovered} with the right time delay, this residual energy is expected to
be much smaller than both $\textrm{SNR}_{\Gamma}$ and the auto-power energy $E_I$
\begin{equation}
  E_I  \equiv \sum \limits_{(t,f) \in \Gamma}|\tilde{y}_I(t;f)|^2
\end{equation}
On the contrary, for a cluster due to a noise outlier in one of the detectors $E^{res}_I$ may become large in the second detector. We can then define a second discriminant variable in addition to $\textrm{SNR}_{\Gamma}$,
\begin{equation}
\Sigma_{res} \equiv\sum_I E^{res}_I / E_I.
\end{equation}
Finally, we combine these two variables into a single detection statistic $\Lambda$ defined by
\begin{equation}
  \Lambda \equiv \frac{\textrm{SNR}_{\Gamma}}{\textrm{SNR}_{\Gamma} + \Sigma_{res}}.
\end{equation}
$\Lambda$ should tend to 1 in presence of a coherent GW signal and take $<< 1$ values in case of noise outliers.
For convenience we define
\begin{equation}
  p_{\Lambda} \equiv -\log(|1-\Lambda|)
\end{equation}
such that the detection statistic increases with the significance of the trigger. Note that $p_{\Lambda}$ is not the only possibility to combine $\textrm{SNR}_{\Gamma}$ and $\Sigma_{res}$. We show in Section \ref{subsec:RD} that $p_{\Lambda}$ is robust to loud noise triggers using a sample of real data from GW detectors, but other combinations may be relevant depending on the distribution of background noise.

\section{Details of the pipeline implementation}
\label{sec:pipeline}
In the following sections, we describe the implementation of \texttt{PySTAMPAS} in the case of a 2-detectors network, and we propose a generalization to the case of network of more than 2 detectors. In practice, the pipeline is implemented using Python 3 and relies on the \texttt{GWpy} package \citep{gwpy}. 

\subsection{Data conditioning}
\label{sec:conditioning}
The GW detectors' data streams are first searched individually to reveal clusters of energy which may contain coherent GW signals. Real GW detector data are available as an ensemble of time-series of different lengths. For a given pair of GW detectors, only coincident times are analyzed. This reduces the data set to a list of \answer{coincident segments of time}. For each of the \answer{coincident segments}, the data are split into windows of duration $T_{win}$ that overlap by 50\%. The duration of the data window is a free parameter that can be adjusted to the typical duration of the GW signal that is being investigated. \answer{In this study, we use $T_{win} \simeq \unit[500]{s}$, as it is done in previous long-duration searches \cite{Abbott:2017muc, Abbott:2019heg}. \texttt{STAMP} was originally designed to search for signals with duration up to several weeks \cite{Thrane:2010ri}. Although there is no fundamental limitation to extending \texttt{PySTAMPAS} to longer signals, we limit ourselves to signals in the range $\unit[10-10^3]{s}$ in this paper.} Working with very large windows leads to dropping up to $T_{win}/2$ s of data at the end of each coincident segments, \answer{and increases the computing cost of clustering.} 

The data are first high-pass filtered to suppress energy outside the analysis frequency bandwidth whose lower boundary is adapted to the GW detectors' noise spectrum of each data set. Real GW detector data often contain non-Gaussian, short duration spikes ("glitches")~\cite{Christensen_2010,Zevin:2016qwy}. When the magnitude of the glitch is large, an excess of energy is present in the $ft$-maps and generates single-detector clusters with very large energy (orders of magnitude larger than what a real GW signal would generate). The coherent step is usually not able to eliminate them completely and a better strategy consists in gating the data time-series before computing the $ft$-maps. \texttt{PySTAMPAS} is mitigating the effect of the loud glitches by applying a Planck window on the $h_I(t)$ samples that exceeds a fixed threshold\footnote{These samples are found by the scipy function \texttt{find\_peaks}.}. \answer{This threshold is a free parameter that should be tuned for each analysis in order to remove most of the glitches without penalizing signal recovery.} After this pre-processing step, $ft$-maps of $\tilde{y}_I(t;f)$ are built.

As shown in all GW detectors' noise spectra ~\cite{Abbott:2019hgc,Buikema:2020dlj}, real GW data contain many spectral artefacts corresponding to mechanical resonances, power lines and pump or fan-like machines surrounding the detectors~\cite{Davis:2021ecd,  LSC:2018vzm}. Most of these spectral lines are of low amplitude and relatively constant over time, while some have a time-varying frequency. \answer{These artifacts can} generate false long duration tracks in $ft$-maps. To attenuate the impact of lines, we set to zero (``frequency notch'') $\tilde{y}_I(t;f)$ pixels corresponding to a list of frequencies that are constructed following two steps:
\begin{enumerate}
    \item For each $ft$-map built, we compute the median value $\bar{y}(f)$ over time of $|\tilde{y}(t;f)|$. Frequencies for which $\bar{y}(f)$ is higher than a given threshold are flagged.
    \item  If a frequency is flagged in more than a given fraction of the total $ft$-maps, it is added to the list.
\end{enumerate}
 This last condition reduces the risk that a monochromatic GW transient \answer{of duration $\lesssim T_{win}$} is mistaken for an instrumental line and notched. \answer{One should note, however, that a very long monochromatic transient signal (on the order of weeks or months) could still be flagged if it is spread over a fraction of the total $ft$-maps higher than the threshold chosen.} 
 If a signal crosses a notched frequency, it may be divided into several parts which will be reconstructed by \texttt{burstegard} as separate clusters, reducing the significance of the signal. To reconnect these parts, we implement the \texttt{findtrack} algorithm \citep{Prestegard:2016}. If the minimal distance between the corners of two clusters is smaller than a given radius, these clusters are connected and treated as one single cluster.

\subsection{Coincident search}
The coincident search is the proper analysis during which true GW signals are searched in the data. The individual detector's $ft$-maps are searched for clusters of excess energy following the procedure described in section \ref{sec:single}. Two lists of clusters are extracted from a pair of detectors. They are saved along with the $ft$-maps to be analyzed in the coherent stage following the procedure described in section \ref{sec:coherent}. The pipeline produces a list of coherent triggers that are ranked according to $p_{\Lambda}$.

\subsection{Background estimation}
\label{subsec:bkg}
In order to assess the significance of triggers found in coincidence, one has to estimate the accidental rate \answer{of noise triggers caused by instrumental and environmental effects}. Like almost all GW transient search pipelines, to encompass any particular effect in the data and augment the total volume of data, we use the time-slides technique to estimate our background~\cite{Was:2009vh}. One data stream is time-shifted with respect to the other one by an amount of time greater than the light travelling time between the detectors. Assuming the number of detectable GW signals is small, this assures that the cross-correlated data does not contain a coherent GW signal. In the meantime, non-Gaussian and non-stationary features of the detectors' noise are preserved. By repeating the analysis for many time-shift values one simulates multiple instances of the noise.

In \texttt{PySTAMPAS}, time-shifts are performed considering data streams split over $N_{win}$ windows that are time ordered on a circle. Data are shifted by a multiple of windows (lags) and for each lag by a multiple of $\Delta t_{max}$ the maximal time resolution (mini-lags). For example, considering only lags, at the \textit{n-th} lag, clusters from detector $I$ that have been extracted in window $i$ are matched with detector $J$ data from window $(i + n)$. With this technique, the maximal number of time-shifts is
\begin{equation}
  (N_{win} - 1) \times \frac{T_{win}}{\Delta t_{max}}.
\end{equation}
The total background lifetime simulated $T_{bkg}$ is the number of time-shifts performed times the duration of data available for a pair of detectors. The cumulative background trigger rate gives an estimation of the false-alarm rate (FAR) as a function of the detection statistic which is used to rank the triggers.

\subsection{Sensitivity studies}
\label{subsec:injections}
\texttt{PySTAMPAS} performs sensitivity studies by injecting simulated signals into the data. A simulated signal consists primarily of a \textit{waveform} which describes the two polarizations modes $h_+(t)$ and $h_{\times}(t)$ of a GW. Waveforms are stored in files in the form of two time series sampled at $f_s$, as well as metadata (duration, frequency range, physical model, etc). A bank of waveforms with various properties is available to \answer{sample the rather large parameter space of long-duration transient GW signals with representative signal morphologies}. 

To compute the detector's response $h_I(t)$ to a given GW signal one has to specify a waveform and the following parameters:
\begin{itemize}
\item the time of arrival $t_0$ at the center of the Earth;
\item the direction $\hat{\Omega}$ to the source;
\item the inclination and polarization angles $(\iota, \, \psi$) that characterize the orientation of the source's reference frame with respect to the Earth equatorial frame;
\item a scaling amplitude factor $\alpha$ to modulate the strength of the signal.
\end{itemize}
Source frame GW polarizations are then rotated to be expressed in the Earth equatorial frame 
\begin{align}
h'_+(t) = a_+ \cos{2\psi}\, h_+(t) - a_{\times}  \sin 2 \psi  \, h_{\times}(t) \notag\\
h'_{\times}(t) = a_+ \sin 2\psi \, h_+(t) + a_{\times}  \cos 2 \psi \, h_{\times}(t)
\end{align}
where $a_+ \equiv \frac{1 + \cos{\iota}^2}{2}$ and $a_{\times} \equiv \cos \iota$\footnote{The dependence of $a_+$ and $a_\times$ on iota is correct for quadrupolar emission.}. The polarizations are then time-shifted by the delay of arrival between the detector's position $\vec{r}_I$ and the center of the Earth
\begin{equation}
\tau_I = \frac{\hat{\Omega} \cdot \vec{r}_I}{c},
\end{equation}
and rescaled by the amplitude factor $\alpha$ such that finally,
\begin{equation}
h_I(t) = \alpha \, [F_I^+(t; \hat{\Omega}) \, h'_+(t - \tau_I) + F_I^{\times}(t; \hat{\Omega}) \, h'_{\times}(t - \tau_I)]
\end{equation}
where $F_I^+(t; \hat{\Omega})$ and $F_I^{\times}(t; \hat{\Omega})$ are the detector's sensitivity to $+$ and $\times$ polarizations (expressed in the Earth equatorial frame) of a GW signal coming from direction $\hat{\Omega}$ at time $t$.
The computed response is resampled and interpolated to match with the detector's sampling, and the first and last seconds of the time series are tapered with a Hann window to avoid numerical artifacts when the signal starts or stops abruptly. 
The signal is injected in the data, which are then analyzed the same way as in a coincident search (restricted to the windows that overlap the injection to gain time). An injection is considered detected if the search produces a trigger within the time and frequency boundaries of the simulated signal, and with a detection statistic $p_{\Lambda}$ larger than a given threshold.

To estimate the detection sensitivity to a given waveform at a given amplitude, a statistically significant number of injections are performed with random starting time, sky position, polarization angle and cosine of the inclination. Starting times are selected in such a way that they always fall within a coincident data segment.
By computing the fraction of recovered injections for different signal amplitudes, it is possible to characterize the detection efficiency as a function of signal's strength, which is usually expressed with the root-sum-squared amplitude $h_{\rm{rss}}$ given by
\begin{equation}
h_{\rm{rss}} \equiv \sqrt{\int (h^2_+(t) + h^2_{\times}(t))\,dt}.
\end{equation}

%

\subsection{Generalization to a network of detectors}
The search algorithm can be generalized in a straightforward manner to a network of $N$ detectors $(I,J,K...)$, constituting $p(N)=N(N-1)/2$ pairs. For a given time-frequency pixel $(t;f)$ and sky direction $\hat{\Omega}$, we define the total coherent SNR as the sum of cross-correlated SNRs from all detector pairs:
\begin{equation}
        \textrm{SNR}(t;f, \hat{\Omega}) = \sum \limits_{I=1}^N \sum \limits_{J>I} \rm{\textrm{SNR}^{IJ}(t;f, \hat{\Omega})},
\end{equation}
with $\rm{\textrm{SNR}^{IJ}(t;f, \hat{\Omega})}$ the coherent SNR computed from Eq.~(\ref{eq:SNR}) corresponding to the pair $IJ$.
This allows us to generalize the definitions of $E_I^{res}$ and $\textrm{SNR}_{\Gamma}$ for a cluster of pixels $\Gamma$,
\begin{align}
  E^{res}_{I} &\equiv \sum \limits_{(t,f) \in \Gamma} |\textrm{SNR}(t;f) / p(N) - |\tilde{y}_I(t;f)|^2|, \\
  \textrm{SNR}_{\Gamma} &\ \equiv\sum \limits_{(t,f) \in \Gamma} \textrm{SNR}(t;f),
\end{align}
and finally the definition of $\Lambda$ remains unchanged:
 \begin{align}
\Lambda \equiv \frac{\textrm{SNR}_{\Gamma}}{\textrm{SNR}_{\Gamma} + \sum \limits_I \frac{E_I^{res}}{E_I}}
 \end{align}

The pipeline's implementation does not fundamentally change with $N \geq 3$ detectors. The clustering step is performed independently over each individual detector's $ft$-maps, \answer{following the hierarchical method of \cite{Thrane:2015psa}} . For each cluster, cross-correlation is computed for the $p(N)$ pairs to compute its ranking statistic $p_{\Lambda}$. However, as the degeneracy between sky direction and time delay between detectors is broken for $N \geq 3$, it is necessary in this case to test all sky positions by choosing uniformly $\alpha$ and $\cos(\delta)$ and select that position that maximizes $\textrm{SNR}_{\Gamma}$. \answer{Therefore, a full-scale study of the pipeline's performances over a network of $3$ or $4$ detectors will be necessary in the future, considering realistic detector's sensitivity curves.}

\section{Performances and comparisons}
\label{sec:performances}

To test the pipeline and demonstrate its performance, we consider $13$ waveforms commonly used in long-duration searches~\cite{Abbott:2017muc,Abbott:2019heg} whose main characteristics are listed in Table~\ref{waveforms}. Most of the waveforms are based on astrophysical models and fall into three categories: eccentric inspiral-merger-ringdown nonspinning compact binary coalescence (ECBC)~\cite{HuMo2017}, broadband chirps from innermost stable circular orbit waves around rotating black holes (ISCOchirp)~\cite{VP2008, VP2016} and accretion disk instability models (ADI)~\cite{vanPutten:2001sw}. We include two \textit{ad hoc} waveforms to better cover the parameter space; a \unit[250]{s} long sine Gaussian signal (SG-C) with a decay time of \unit[50]{s} and a \unit[20]{s} long band-limited white noise burst (WNB-A). These signals of different morphology cover the time-frequency space with durations within \unit[9-290]{s} and frequencies in the \unit[10-2048]{Hz} range. In the following we consider the case of a 2-detector search to compare performance with \texttt{STAMP-AS}. If not stated differently, we are using simulated Gaussian noise following LIGO's best sensitivity during the second observing run (O2)~\citep{Abbott_2019} to simulate the data from the two LIGO detectors at Hanford (H1) \cite{O2_H1_sens} and Livingston (L1) \cite{O2_L1_sens}.
\begin{table*}[!htb]
\footnotesize
\centering
\begin{tabular}{lcccc}
\hline
Waveform & Parameters & Duration [s] & Frequency [Hz]& Morphology\\
\hline
ECBC-A & $M_1=1.4~M_{\odot}$, $M_2=1.4~M_{\odot}$, $ecc=0.2$& 291 & 10 - 250  & Chirp\\
ECBC-B & $M_1=1.4~M_{\odot}$, $M_2=1.4~M_{\odot}$, $ecc=0.4$ & 178 & 10 - 275  & -\\
ECBC-C & $M_1=1.4~M_{\odot}$, $M_2=1.4~M_{\odot}$, $ecc=0.6$ & 64 & 10 - 350 & -\\
ECBC-D & $M_1=3.0~M_{\odot}$, $M_2=3.0~M_{\odot}$, $ecc=0.2$ & 81 & 10 - 180 &  -\\
ECBC-E & $M_1=3.0~M_{\odot}$, $M_2=3.0~M_{\odot}$, $ecc=0.4$ & 49 & 10 - 200 &-\\
ECBC-F & $M_1=3.0~M_{\odot}$, $M_2=3.0~M_{\odot}$, $ecc=0.6$ & 15 & 10 - 200 & -\\
ISCOchirp-A & $m_{BH}= 5.0~M_{\odot}$ & 237 & 1049 - 2048 &  Broadband chirp-down\\
ISCOchirp-B & $m_{BH}= 10.0~M_{\odot}$& 237 & 705 - 2048 &  -\\
ISCOchirp-C & $m_{BH}=20.0~M_{\odot}$ & 236 & 196 - 1545 &  -\\
ADI-A & $m_{BH}= 5.0~M_{\odot}$, $a_{BH}=0.3$ & 35 & 135 - 166 & Chirp-down\\
ADI-B & $m_{BH}= 10.0~M_{\odot}$, $a_{BH}=0.95$ & 9 & 110 - 209 & -\\
SG-C &  & $243$ & $402$-$408$ & Mono-chromatic \\
WNB-A & & 20 & 50-400 &  Band limited white noise \\
\hline
\end{tabular}
\caption{Name, parameters, duration, frequency range and spectral morphology of waveforms used to characterize \texttt{PySTAMPAS}. $M_i$ is the component compact object mass; $ecc$ is the eccentricity of the binary orbit at \unit[10]{Hz}; $M_{BH}$ and $a_{BH}$ are the mass and normalized spin of the black hole.}
\label{waveforms}
\end{table*}

\subsection{Signal reconstruction}
We investigate the effects of several parameters of the pipeline on the detection capability and the signal reconstruction in order to find a set of parameters that maximize the detection of a wide range of different morphology signals, while keeping the computational costs affordable.
\subsubsection{Power spectral density estimation}
\label{subsec:PSD}
The accuracy of the noise power spectral density (PSD) estimation is playing a central role to reconstruct GW signals efficiently. Yet, this task is complicated in the case of GW detectors as the noise contains non-Gaussian and non-stationary features such as glitches, spectral lines and slow drifts of the noise amplitude.

Consider a detector's strain time series given by $s_I(t) = h_I(t) + n_I(t)$, where $h_I(t)$ is a deterministic GW signal and $n_I(t)$ random noise. A good estimator of the one-sided PSD of the noise is given by the squared modulus of its Fourier transform,
\begin{equation}
P_I(t;f) \equiv \langle|\tilde{n_I}(t;f)|^2 \rangle,
\end{equation}
This value is not directly accessible because in case of an unknown GW waveform it is not possible to disentangle \textit{a priori} signal from noise. One has to rely on the observable $|\tilde{s_I}(t;f)|^2$, that may contain GW signal. Assuming signal and noise are not correlated,
\begin{equation}
\langle\,|\tilde{s_I}(t;f)|^2 \,\rangle =   \langle\,|\tilde{h_I}(t;f)|^2 \,\rangle + P_I(t;f).
\end{equation}
Therefore, an assumption over the nature of the signal $\tilde{h}_I(t;f)$ must be made in order to build an unbiased estimator of $P_I(t;f)$.  \texttt{PySTAMPAS} implements two methods to estimate the PSD that are suited for different signal morphologies.

The first method consists in taking the average of $|\tilde{s}_I(t;f)|^2$ over $n_t$ symmetrically chosen neighbouring \answer{Fourier transformed segments}. The underlying assumptions are that (1) the noise is stationary over the time window considered, and (2) no signal is present in the adjacent pixels. As discussed above, (1) is often wrong because of the presence of short glitches in the data, which are therefore not factored in the PSD and appear as signal. Conversely, (2) is wrong when a monochromatic or quasi-monochromatic signal is present in the data, leading these to be mistakenly included in the PSD. Degraded sensitivity to monochromatic signals is a known weakness of \texttt{STAMP} \cite{Abbott:2015vir}. 

\begin{figure}
    \centering
    \includegraphics[width=\columnwidth]{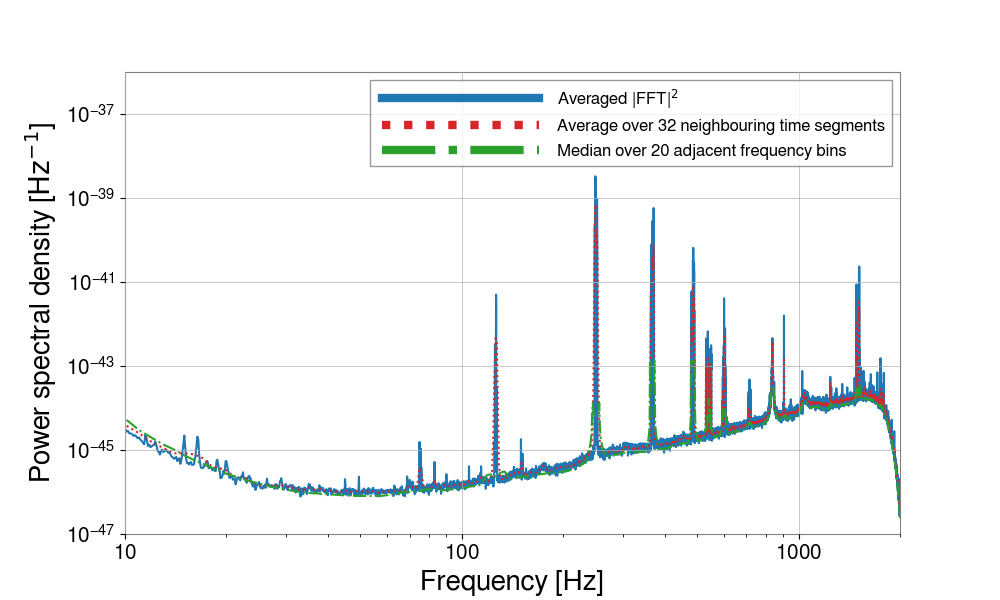}
    \caption{Estimation of the PSD for a \unit[100]{s} long segment of LIGO Hanford data from the O2 observing run using the two different methods implemented in \texttt{PySTAMPAS}. The squared modulus of the Fourier transform (\answer{averaged over 10 independent realizations of the noise}) is shown in blue for reference.}
    \label{fig:psd_comparison}
\end{figure}

To address these issues, we propose to estimate the PSD by taking the median of $|\tilde{s}_I(t;f)|^2$ over $n_f$ adjacent frequency bins. The pros and cons of this method are opposite to the first one: short glitches are well taken into account and monochromatic signals are better reconstructed. However, signals whose frequency evolution is rapid tend to be less well reconstructed. In case of noise only, both methods provide similar PSD estimates, except that spectral narrow features are better reconstructed with the method averaging the neighboring time-segments pixels, as shown in Fig.~\ref{fig:psd_comparison}. We use the median as it is more robust that the average to extreme values. Because of instrumental lines, it is likely that one of the neighbouring frequency bins has pixels with a very high value of s(t;f), which would spoil the PSD estimation.

\begin{figure*}
\includegraphics[width=\columnwidth]{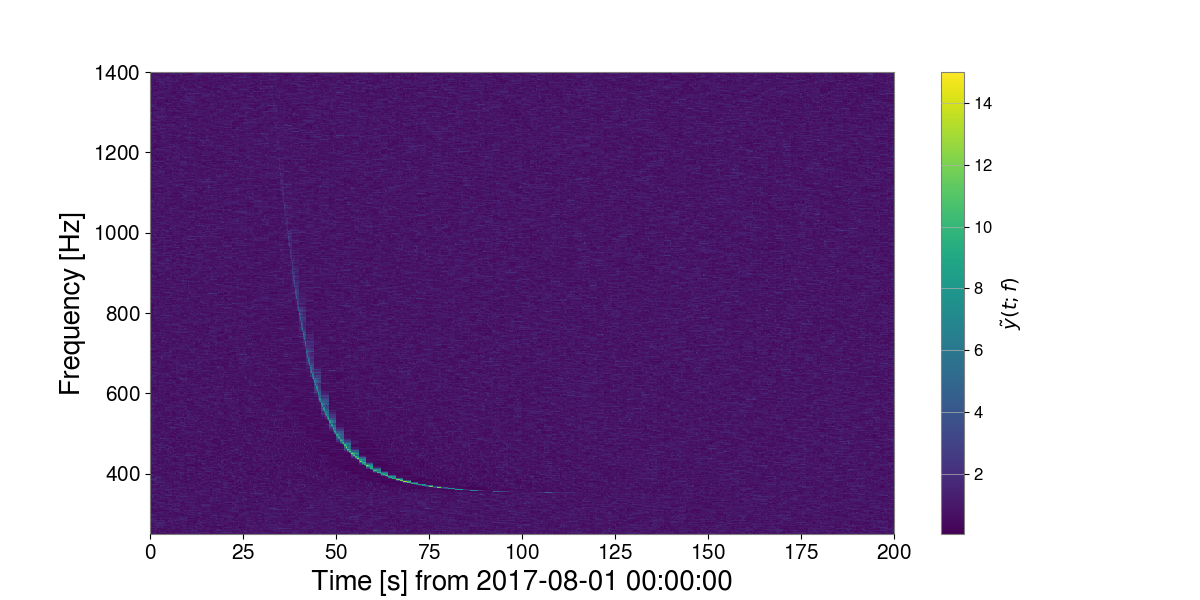}
\includegraphics[width=\columnwidth]{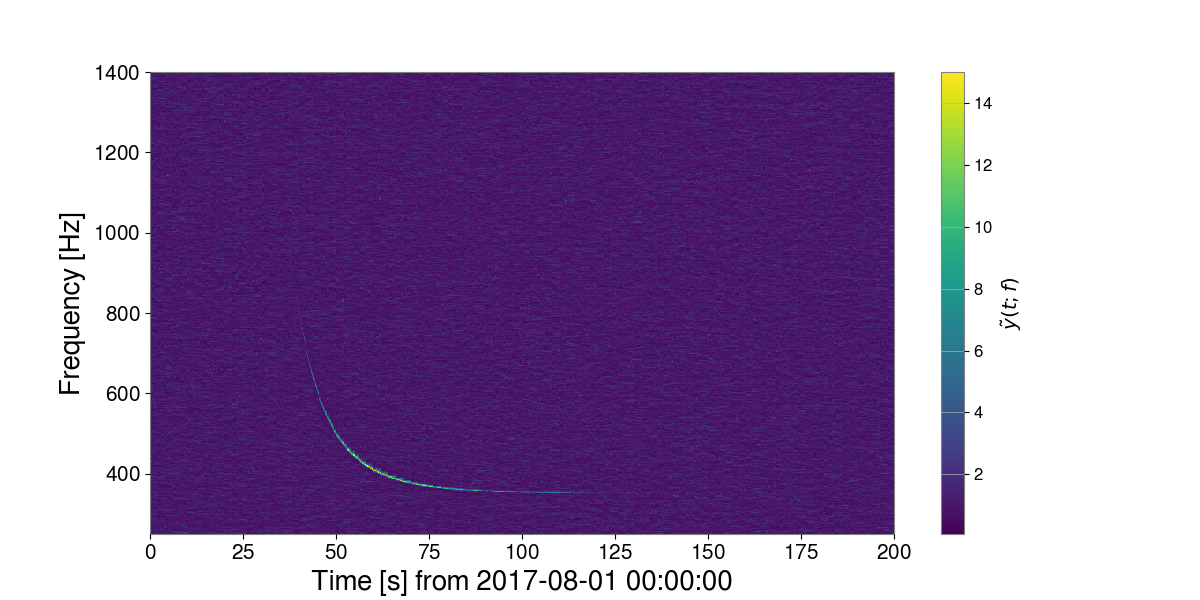}
\includegraphics[width=\columnwidth]{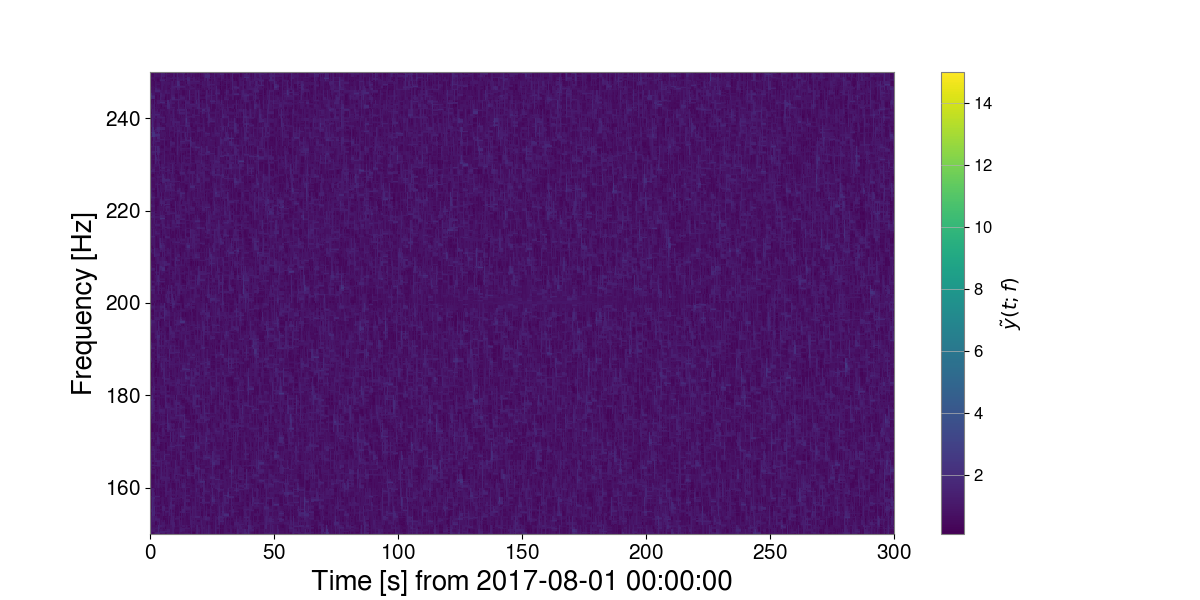}
\includegraphics[width=\columnwidth]{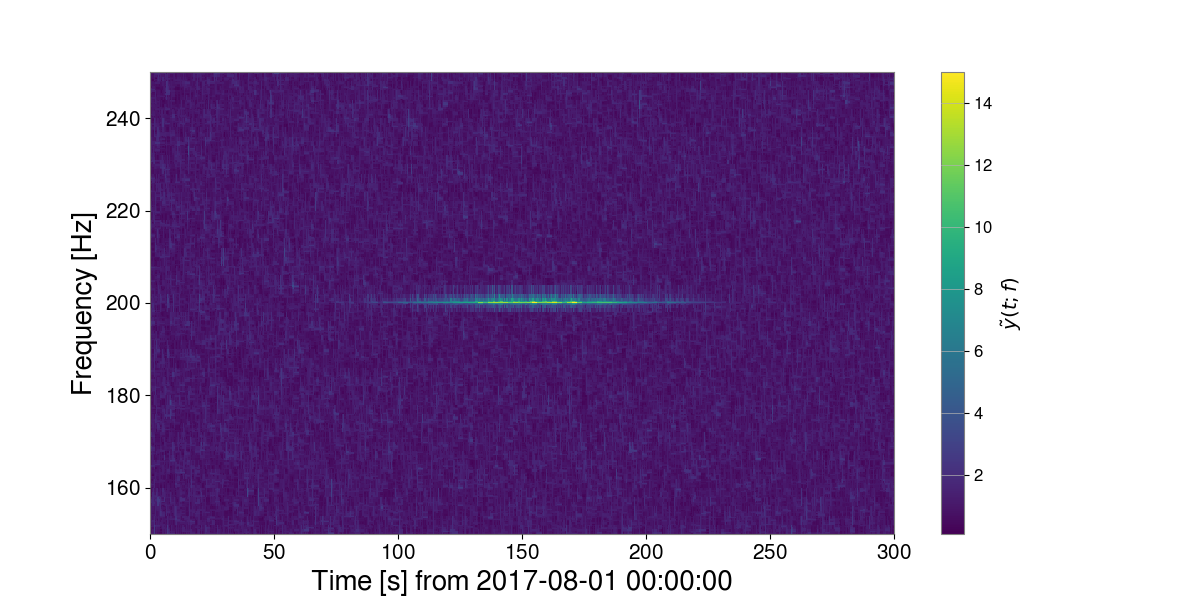}
\caption{Time-frequency maps of |$\tilde{y}(t;f)$| for two injected signals (\textit{top}: ISCOchirp, \textit{bottom}: sine Gaussian) realised with the two different PSD methods (\textit{left}: average over $n=32$ adjacent time bins (\textit{time-average}), \textit{right}: moving median over $n=20$ frequency bins (\textit{frequency-median})).}
\label{ftmaps}
\end{figure*}

The effect of the PSD estimation on the signal reconstruction in \text{PySTAMPAS} is illustrated in Fig.~\ref{ftmaps}. Two signals with very different spectral morphologies, a broadband ISCO chirp (ISCOchirp-C) \cite{VP2016} and a monochromatic sine Gaussian (SG-C), are injected in Gaussian noise. By taking the median over adjacent frequency bins (hereafter referred as \textit{frequency-median PSD}) instead of averaging over neighbouring \answer{Fourier transformed segments} (\textit{time-average PSD}), the sine Gaussian signal is better reconstructed, but the fast frequency evoluting part of the ISCOchirp is blurred out. The optimal choice of a PSD estimation method depends on the type of signals targeted and the characteristics of the noise, especially spectral lines and/or non-stationary features. Another way to restore the sensitivity to monochromatic triggers would be to consider a very long (\unit[$\sim 10^3$]{s}) time period to estimate the PSD in the case of the time adjacent pixels method. However, \answer{noise from GW detectors tend to become non-stationary over such time intervals at low frequencies (below $\sim 100$ Hz) \cite{Davis:2021ecd}}.

\subsubsection{Source sky location determination}
\label{subsec:skylocation}
The number of sky locations tested in the coherent step (which reduces to a single time delay parameter in the case of a two detector network) is currently a limiting factor of all-sky searches, and illustrates the necessary trade-off between detection sensitivity and computational cost \cite{Abbott:2015vir}. The hierarchical processing implemented in \texttt{PySTAMPAS} allows for scanning many positions at a low cost. In Section \ref{sec:coherent}, we have seen that $N_\tau$, the number of time delays between detectors to be tested, depends on the ratio between the maximal and the minimal frequency of the trigger. Here, we investigate empirically the pipeline sensitivity loss as function of the number of time delays for different waveform families.

 \answer{Signal waveforms} are injected coherently into Gaussian noise, \answer{simulating} data from LIGO Hanford and LIGO Livingston, from a given sky direction $\hat{\Omega}_0$, and are \answer{recovered} by \texttt{PySTAMPAS}. We vary the number of time delays and keep the maximal $\textrm{SNR}_{\Gamma}$ obtained which is compared to $\textrm{SNR}_{\Gamma}(\hat{\Omega}_0)$, the SNR value corresponding to the true source position $\hat{\Omega}_0$. 

\begin{figure}
\includegraphics[width=\columnwidth]{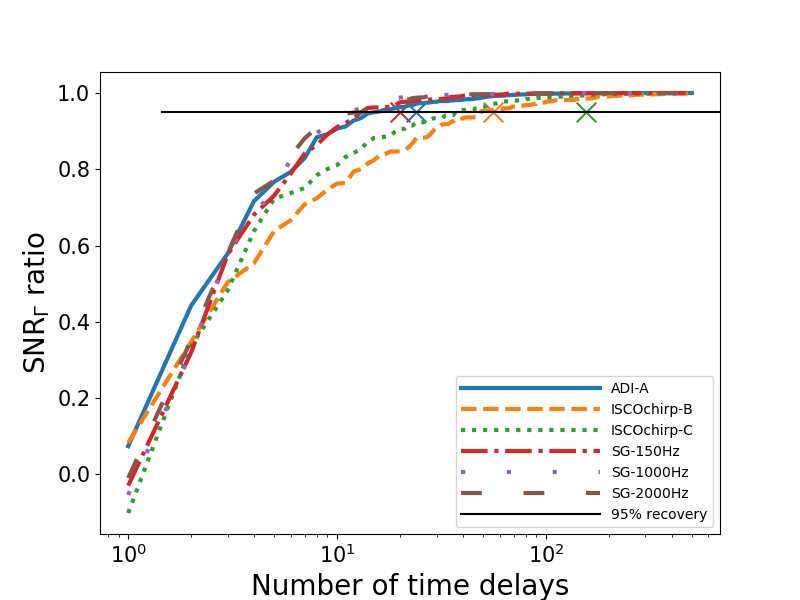}
\caption{Ratio between the \answer{recovered} $\textrm{SNR}_{\Gamma}$ and $\textrm{SNR}_{\Gamma}(\hat{\Omega_0})$ as function of the number of time delays between the LIGO Hanford and LIGO Livingston detectors for different waveforms. $50$ injections have been performed for each value of $N_{\tau}$. Crosses represent the theoretical number of time delays to test to reach a ratio of $0.95$, computed from Eq. \ref{eq:ntau}.}
\label{fig:snrloss}
\end{figure}

The ratio $\textrm{SNR}_{\Gamma}$ to $\textrm{SNR}_{\Gamma}(\hat{\Omega}_0)$ as function of the number of time delays between detectors is shown in Fig.~\ref{fig:snrloss} for sine Gaussians of different central frequency and for a selection of waveforms of different morphology/durations.
We compare the number $N_{\tau}$ of delays tested to get $\epsilon = 0.95$ to the theoretical prediction \answer{from Eq. (\ref{eq:ntau})} given in Table~\ref{tab:n_tau}.
We see that the optimal value of $N_\tau$ does not depend on the signal frequency, but mainly on its frequency range $f_{max} / f_{min}$. Monochromatic sine Gaussians are \answer{recovered} equally rapidly no matter their frequency, and faster than signals of broader band.
\answer{The empirical values are overall lower than the theoretical ones. This discrepancy comes from the fact that the clustering algorithm does not always reconstruct the entirety of the waveform, leading to a lower effective value of $f_{max} / f_{min}$.}
To optimize the \answer{detection} efficiency while keeping the number of tested sky positions minimal, we fix $\epsilon$ such that the maximal SNR loss parameter to 5\% and $N_\tau$ is determined for each cluster following Eq.~(\ref{eq:ntau}).

\begin{table}
    \centering
\begin{tabular}{lccc}
\hline
 Waveform   &   $f_{max}/f_{min}$ &   $N_{\tau}$ \\
\hline
 ADI-A       &         1.2 &     24 \\
 ADI-B       &         1.9 &     37 \\
 ISCOchirp-A &         1.9 &     37 \\
 ISCOchirp-B &         2.8 &     56 \\
 ISCOchirp-C &         7.9 &    155 \\
 ECBC-A  &        12.5 &    247 \\
 ECBC-B  &        13.8 &    272 \\
 ECBC-C  &        17.5 &    346 \\
 ECBC-D  &         9   &    178 \\
 ECBC-E  &        10   &    197 \\
 ECBC-F  &        10   &    197 \\
 SG-C        &         1   &     20 \\
 WNB-A       &         1.2 &     23 \\
\hline
\end{tabular}    \caption{Theoretical minimal number $N_{\tau}$ of time delays between two detectors (here LIGO Hanford and LIGO Livingston) to be considered for each waveform in order to \answer{recover} the coherent signal SNR with an accuracy larger than 0.95. $N_{\tau}$ depends on the frequency ratio $f_{max} / f_{min}$ of the signal considered as given in Eq.~(\ref{eq:ntau}).}
\label{tab:n_tau}
\end{table}

\subsubsection{Multi-resolution and clustering}
The energy of long-duration GW signals is spread over a potentially large number of pixels. This would mean it is necessary to lower the threshold on the individual pixel's energy $|\tilde{y}_I(t;f)|$ and rely on the clustering algorithm to group all pixels belonging to the cluster. Clustering a large number $N$ of pixels is computationally expensive since \texttt{burstegard}'s time complexity is $\mathcal{O}( N \log N)$. However, because of the hierarchical implementation, that step is computed only once per $ft$-map and is therefore no longer a bottleneck for analyzing long periods of data. Yet, the risk is to include pixels due to noise fluctuations and generate clusters that are only composed of noise pixels. By increasing the minimal number of pixels per cluster, one can control the rate of noise clusters that are generated.  

Another way to collect, as best as possible, all the energy in the $ft$-maps is to process the data with a range of different time-frequency resolutions that match well all the different GW signal shapes. 
The choice of time-frequency resolutions depends on the waveform, but we have seen that for the diversity of signals we are targeting, a limited number of time-resolutions is enough to improve the detection efficiency of non-monochromatic GW signals. Using a set of $4$ resolutions ranging from \unit[4]{s} $\times$ \unit[0.25]{Hz} to  \unit[0.5]{s} $\times$ \unit[2]{Hz}, we report an efficiency increase by $5-40 \%$ for the waveforms tested (at constant FAR), compared to \unit[1]{s} $\times$ \unit[1]{Hz} pixels. \answer{We report the relative increase in detection efficiency for each astrophysical waveform in Table \ref{tab:multires}.}

\begin{table}
    \centering
\begin{tabular}{lcc}
\hline
 Waveform   &  Effiency increase \\
\hline
 ADI-A       &         $+25\%$\\
 ADI-B       &         $+5\%$ \\
 ISCOchirp-A &         $+17\%$\\
 ISCOchirp-B &         $+10\%$ \\
 ISCOchirp-C &         $+18\%$ \\
 ECBC-A  &        $+23\%$ & \\
 ECBC-B  &        $+17\%$ &  \\
 ECBC-C  &        $+40\%$ & \\
 ECBC-D  &         $+38\%$   &\\
 ECBC-E  &       $+41\%$   & \\
 ECBC-F  &       $+3\%$   & \\
\hline
\end{tabular}    
\caption{Relative increase on the distance at $50\%$ detection efficiency between a single-resolution approach (pixels of $\unit[1]{s} \times \unit[1]{Hz}$) and the multi-resolution approach implemented in \texttt{PySTAMPAS} (4 different resolutions from $\unit[0.5]{s} \times \unit[2]{Hz}$ to $\unit[4]{s} \times \unit[0.25]{Hz}$), all other parameters being equal, for the astrophysical waveforms described in Table \ref{tab:parameters}}
\label{tab:multires}
\end{table}

It is not possible to perform a fine optimization of all \texttt{PySTAMPAS} parameters for a generic all-sky/all-time search because of the large parameter space, but we present in the next Sections the pipeline performance for both Gaussian simulated noise and real GW data to detect long duration GW signals using the set of parameters given in Table~\ref{tab:parameters}.
\begin{table}[H]
\centering
\begin{tabular}{lc}
\hline
Parameters & Value \\
\hline
\textbf{$ft$-maps} & \\
Window duration & $512s$ \\
Frequency range & $20-2000$ Hz \\
\multirow{2}{*}{$\Delta t_i \times \Delta f_i$} & $[ 4.0 \, \textrm{s} \times 0.25 \, \textrm{Hz} - 2.0 \, \textrm{s}  \times 0.5 \, \textrm{Hz}$ \\
& $- 1.0 \, \textrm{s}  \times 1.0 \, \textrm{Hz}- 0.5 \, \textrm{s}  \times 2.0 \, \textrm{Hz}]$\\
\hline
\textbf{PSD estimation} & \\
Time-average & $32$ time bins \\
Frequency-median & \unit[20]{Hz}\\
\hline
\textbf{Clustering} & \\
Pixel energy threshold & $2.0$ \\
Clustering radius & $\unit[2]{s} \times \unit[2]{Hz}$\\
Minimum pixels number & $30$ \\
\hline
\textbf{Coherent stage} & \\
SNR loss $1-\epsilon$ & 5\% \\

\hline
\end{tabular}
\caption{\texttt{PySTAMPAS} parameter values used in the all-sky/all-time long-duration GW search with Advanced LIGO / Advanced Virgo data presented in this paper.}
\label{tab:parameters}
\end{table}

\subsection{Test on simulated data}

We carry out a study with simulated Gaussian noise to test the pipeline as a whole and evaluate its performance. First, we generate two sets of $14$ days of stationary Gaussian GW noise following LIGO's O2 sensitivity to simulate the data from the two LIGO detectors at Hanford and Livingston. We analyze these data with \texttt{PySTAMPAS}, using parameters given in Table \ref{tab:parameters}.

Background triggers are generated following the method described in \ref{subsec:bkg}. We perform $128,000$ time-slides, simulating $\sim 4,900$ years of background noise accounting for $34$ days of CPU time on a dual-core modern processor. As a comparison, the previous version of \texttt{STAMP-AS} took $95$ days of CPU time to perform $1,000$ time-slides over the same data, meaning that \texttt{PySTAMPAS} is faster by at least one order of magnitude. On Fig.~\ref{fig:far_rd} showing the cumulative false alarm rate (FAR) as function of $p_{\Lambda}$, the blue curves correspond to the distribution of simulated Gaussian noise triggers for the two PSD estimation methods; the shape of the two curves is similar, but the median-frequency PSD method produces $\sim$ 60 more triggers \answer{than the time-average PSD}. This has little \answer{effect} on the pipeline sensitivity as the tail of the $p_{\Lambda}$ distribution are similar.

\begin{figure}
\includegraphics[width=\columnwidth]{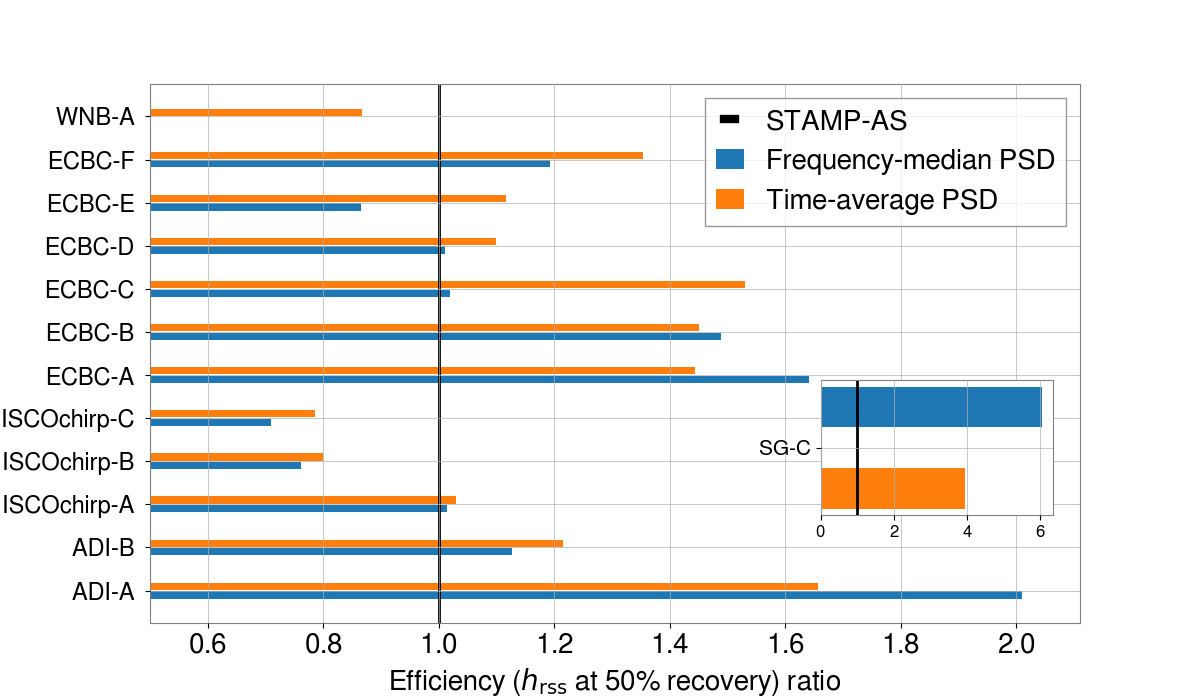}
\caption{Ratio between the $h_{\rm{rss}}$ at $50 \%$ detection efficiency obtained with \texttt{STAMP-AS} and with \texttt{PySTAMPAS} for a $\rm{FAR} = 1/50 \, \rm{yr}^{-1}$ for both PSD estimation methods. The white noise burst waveform WNB-A was not recovered at all using the frequency-median PSD. \answer{A ratio above $1$ means that \texttt{PySTAMPAS} recovers the signal better than \texttt{STAMP-AS}.}}
\label{fig:efficiency_MC}
\end{figure}

For each waveform described in Table \ref{waveforms}, we estimate the detection efficiency as function of $h_{\rm{rss}}$ following the method described in section \ref{subsec:injections}. We fix a detection threshold corresponding to a FAR of $1/50 \,\rm{yr}^{-1}$ and determine the value $h_{\rm{rss}}^{50\%}$ of $h_{\rm{rss}}$ for which $50 \%$ of the injections are recovered. To provide a comparison, we perform the same search with \texttt{STAMP-AS} over the same simulated Gaussian noise. We use the quantity $h_{\rm{rss}}^{50\%}$ to estimate the detection efficiency of the search. It is inversely proportional to the typical detection range.
In Fig.~\ref{fig:efficiency_MC}, we show the ratio of $h_{\rm{rss}}^{50\%}$ between \texttt{STAMP-AS} and \texttt{PySTAMPAS} for each waveform and each PSD estimator.

For a majority of the waveforms tested, \texttt{PySTAMPAS} is more sensitive than \texttt{STAMP-AS}, up to a factor 2, with the exception of the ISCOchirp family for which detection efficiencies are worse by down a factor $0.8-1$ \answer{in the best case with the time-average PSD}. For this specific family, the single-detector clustering algorithm reconstructs low amplitude signals poorly because the energy is spread over too many pixels. \answer{Down to a certain amplitude, most pixels fall below the clustering threshold and the signal is not reconstructed at all. A finer tuning of \texttt{burstegard} could be done to address this limitation, but this type of signal would certainly be better reconstructed by seedless clustering algorithms. This also illustrates the difficulty of tuning the pipeline to maximize sensitivity to a wide variety of waveforms.}

The \textit{ad-hoc} waveforms illustrate the most extreme cases. Detection efficiency is multiplied by $\sim 6$ for the monochromatic sine Gaussian signal (SG-C) when the PSD is computed over adjacent frequency bins as compared to \texttt{STAMP-AS}.  On the other hand, the large band white noise burst (WNB-A) is not \answer{recovered} at all with this method, and \answer{recovered} almost equally well with the time-average PSD. We note that the sine Gaussian is also better \answer{recovered} using the time-average PSD. This is due to the fact that we consider a wider time window to compute the PSD ($32$ pixels from each side instead of $8$).

\subsection{Tests on real data}
\label{subsec:RD}
\begin{figure}[h]
\includegraphics[width=\columnwidth]{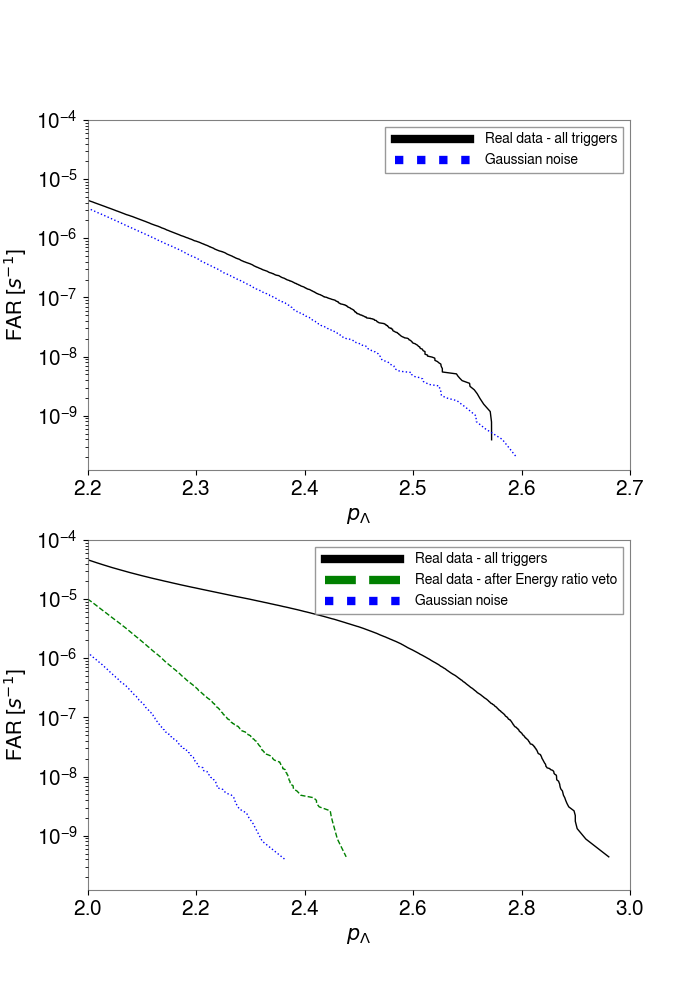}
\caption{FAR obtained with data from LIGO O2 observing run versus the detection statistic $p_{\Lambda}$ with frequency-median PSD (top) and time-average PSD (bottom). The blue curves represent the FAR obtained with Gaussian noise. FAR of triggers remaining after applying Rveto is shown by the green curve.}
\label{fig:far_rd}
\end{figure}

Real data from GW detectors have non-Gaussian and non-stationary features that challenge pipelines. To understand the behaviour of \texttt{PySTAMPAS} on real GW noise, we analyze LIGO data from the Advanced LIGO and Advanced Virgo O2 observing run downloaded from the Gravitational Wave Open Science Center~\citep{Rich_Abbott_2021, O2_DOI}. The chosen period runs from 2017-08-01 00:00:00 UTC to 2017-08-15 00:00:00 UTC and contains $9.21$ days of coincident data from H1 and L1. We keep the pipeline's parameters given in Table \ref{tab:parameters}, but switch on the spectral lines removal algorithm described in Section~\ref{sec:conditioning}. About 5\% of the total frequency bins are flagged as spectral lines and notched for each detector. As in the simulated data study, we consider both PSD estimation methods.
Cumulative FAR distributions for O2 data are compared to simulated Gaussian noise FAR distributions in Fig.~\ref{fig:far_rd}. For both PSD estimation methods, an excess of triggers is present compared to the simulated Gaussian noise distributions, meaning that the FAR of the search for a given value of $p_{\Lambda}$ is higher than with Gaussian noise.

For the frequency-median PSD, the excess of triggers ($\sim 20 \%$ more triggers in real data \answer{than in the Monte Carlo study with Gaussian noise}) consist \answer{of} long-duration ($>$ \unit[50]{s}), quasi-monochromatic events that correspond to instrumental lines being punctually excited. These lines are too low amplitude and are not excited regularly enough to be flagged by the spectral lines removal algorithm. However, that excess becomes marginal for large value of $p_{\Lambda}$ and thus is not affecting the overall pipeline sensitivity for this set of data.

\begin{figure}[h]
\includegraphics[width=\columnwidth]{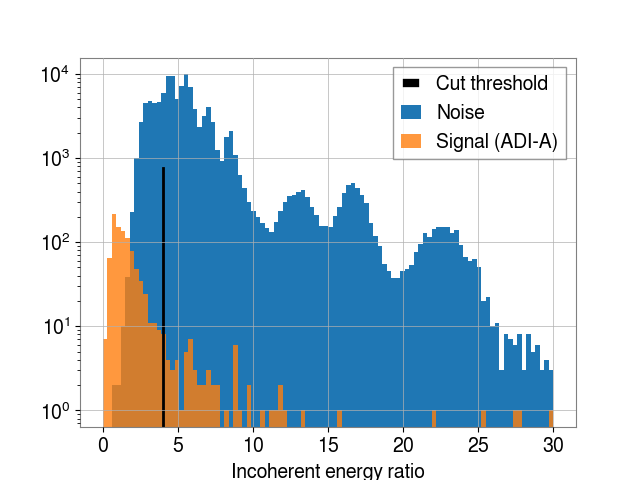}
\caption{Distribution of the incoherent energy ratio $R$ obtained for background triggers (in blue) and GW signal triggers from the ADI-A waveform (in orange) using the time-average PSD. Rejecting triggers with $R>4$ \answer{allows for reducing the excess of large $p_{\Lambda}$ background triggers}, while marginally affecting the pipeline efficiency to recover GW signals.}
\label{fig:Rveto}
\end{figure}

Using the time-average PSD method, the excess of triggers \answer{compared to Gaussian noise} is much larger, by at least $1.5$ orders of magnitude. It is dominated by short glitches with frequencies between \unit[20-100]{Hz} which have passed the gating procedure. They generate triggers with high $p_{\Lambda}$ that populate the tail of the distribution. 
To discriminate those triggers, we implement a veto, \textit{Rveto}, based on the ratio of incoherent energy between the detectors $R = E_I /E_J$, similar to what is done for \texttt{STAMP-AS} in \cite{Abbott:2019heg}. 
Fig. \ref{fig:Rveto} shows the cumulative distributions of $R$ for background triggers and for triggers recovered for a GW waveform (ADI-A). Vetoing triggers with $R>4$ allows to reduce by a factor 5 the number of triggers but more interestingly, the tail of the distribution of $p_{\Lambda}$ is drastically reduced to approach the Gaussian noise triggers estimation, while no more than $5\%$ of GW signal triggers are vetoed.
In this paper, we are just illustrating that the pipeline behavior changes considerably in presence of non Gaussian and non stationary data. We also show that simple post-processing selection criteria can be easily developed and applied \answer{with a relatively small} penalty for the overall pipeline sensitivity.  


As we have done for the study with simulated Gaussian noise, we now estimate the detection efficiency of this search with the two PSD estimators and compare it to results obtained by \texttt{STAMP-AS} during the second Advanced LIGO observing run \cite{Abbott:2019heg} for a FAR of \unit[$1/50$]{$\rm{yr}^{-1}$}. \answer{For the time-average PSD, signals with $R>4$ are rejected like is done in the background study.}
Best results obtained for each waveform among the two PSD methods are presented in Table \ref{tab:RD_efficiency} and compared to \texttt{STAMP-AS}. The relative detection efficiency depends on the waveform, but the overall \texttt{PySTAMPAS} pipeline efficiency increase observed with real data is very similar to what was obtained on simulated Gaussian data.
\begin{table}
    \footnotesize
    \centering
    \begin{tabular}{l|ccc|c}
    \multirow{2}{*}{Waveform}& \multicolumn{2}{c}{\texttt{PySTAMPAS}} & \texttt{STAMP-AS} &  \multirow{2}{*}{Ratio}\\
     & \textit{frequency-median} & \textit{time-average} &  &  \\
    \hline
 ISCOchirp-A & $9.18 \times 10^{-21}$ & $8.17 \times 10^{-21}$  &$6.20 \times 10^{-21}$  & $0.76$  \\
 ISCOchirp-B & $1.84 \times 10^{-21}$ & $2.01 \times 10^{-21}$ & $1.44 \times 10^{-21}$ & $0.78$ \\
 ISCOchirp-C & $8.89 \times 10^{-22}$ & $1.06 \times 10^{-21}$ & $1.01 \times 10^{-21}$ & $0.95$ \\
 ECBC-A  & $9.95 \times 10^{-22}$ & $1.07 \times 10^{-21}$ & $1.55 \times 10^{-21}$ & $1.55$ \\
 ECBC-B  & $8.81 \times 10^{-22}$ & $8.61 \times 10^{-22}$ & $1.34 \times 10^{-21}$ & $1.56$ \\
 ECBC-C  & $8.64 \times 10^{-22}$ & $8.00 \times 10^{-22}$ & $1.35 \times 10^{-21}$ & $1.69$ \\
 ECBC-D  & $1.20 \times 10^{-21}$  & $8.95 \times 10^{-22} $ & $1.48 \times 10^{-21}$ & $1.65$ \\
 ECBC-E  & $1.12 \times 10^{-21}$ & $8.82 \times 10^{-22}$ & $1.89 \times 10^{-21}$ & $2.14$ \\
 ECBC-F  & $9.25 \times 10^{-22}$ & $7.83 \times 10^{-22}$ & $9.64 \times 10^{-22}$ & $1.23$ \\
 ADI-B   & $3.26 \times 10^{-22}$ & $3.26 \times 10^{-22}$ & $4.81 \times 10^{-22}$ & $1.47$ \\ 
 SG-C   & $4.34 \times 10^{-22}$& $6.88 \times 10^{-22}$ & $4.35\times10^{-21}$ & $10.0$ \\ 
 WNB-A  & -& $2.0 \times10^{-21}$ & $2.04\times10^{-21}$ &$1.00$ \\ \end{tabular}    

 \caption{Values of $h_{\rm{rss}}$ at $50\%$ detection efficiency for different waveforms obtained with \texttt{PySTAMPAS} for the two PSD methods and \text{STAMP-AS} over O2 data from LIGO Hanford and LIGO Livingston, using a FAR threshold of \unit[$1/50$]{$\rm{yr}^{-1}$}. The last column shows the ratio between \texttt{STAMP-AS} and the lowest value of \texttt{PySTAMPAS} among the two PSD methods. White noise burst waveforms WNB-A are not recovered at all with the frequency-median PSD.}
    \label{tab:RD_efficiency}
\end{table}

\section{Conclusion}
\label{sec:conclusion}
In this paper, we have presented \texttt{PySTAMPAS}, a new data analysis pipeline designed to search for GW \answer{of duration $\unit[\sim 10 - 10^3]{s}$} in a network of detectors with minimal assumptions on the nature and origin of the signal.
The search algorithm relies on a hierarchical method, initially designed for a seedless clustering algorithm \cite{Thrane:2015psa}, where candidate events are first identified in single-detectors $ft$-maps, and a coherent detection statistic is then computed by cross-correlating data streams from each pair of detector. This method provides a significant gain in computational efficiency compared to the initial implementation of \texttt{STAMP-AS} with seed-based clustering, while still benefiting from the increased sensitivity of coherent searches. This is especially critical for all-sky/all-time searches for which both the data set and the parameter space can be very large.

The reduced computational cost allows us to implement several new features to improve the overall sensitivity of the pipeline. The use of multi-resolution $ft$-maps enables the better reconstruction of signals with fast frequency evolution. An alternative method to estimate the noise PSD is proposed that is best suited for monochromatic and quasi-monochromatic signals. We also introduce a new detection statistic that compares the coherent SNR of an event to the incoherent auto-power in single detectors in order to discriminate coherent GW signals from loud noise events. Additionally, it is now feasible to scan hundreds of sky positions during the coherence stage, and therefore to reduce the loss of SNR due to an error in the sky position to less than $5\%$.
The combination of these features results in a detection efficiency increased by a factor $\sim 1.5$ \answer{on average} compared to the previous version of \texttt{STAMP-AS} with seed-based clustering \answer{for the different waveforms tested, which have durations between $8-291$\,s, frequencies between $10-2048$ Hz, and various spectral morphologies}. \answer{We note that the changes in detection efficiency are dependant on the type of waveform, with \texttt{PySTAMPAS} performing slightly less well on waveforms from the ISCO chirp family and better for the remaining waveforms. We plan to improve the tuning of the clustering algorithm to address this issue.}

\texttt{PySTAMPAS} is able to perform all-sky or targeted searches over a full observing run and a network of detectors, and provides a basis for further developments.
For example, the \texttt{burstegard} algorithm has been used here to identify clusters of excess power pixels, but other detection algorithms could be considered, such as seedless clustering \citep{seedless} or more complex pattern recognition algorithms. \answer{This will be need to be done in order for the pipeline to be fully competitive, as show by the example of the ISCO chirp waveforms family, which are currently slightly less well recovered by \texttt{PySTAMPAS}.}
We have shown that real GW data search requires to develop specific trigger selection to cope with non Gaussian and non stationary features of GW detectors data, but another possibility of improvement could consist in implementing a better identification and subtraction of non-Gaussian features of the GW detectors noise, as well as better discriminant variables.

\bigskip\noindent\textit{Acknowledgments} ---
N.C. is supported by NSF grant PHY-1806990. M.C. is supported by NSF grant PHY-2010970. This research has made use of data, software and/or web tools obtained from the Gravitational Wave Open Science Center (https://www.gw-openscience.org/), a service of LIGO Laboratory, the LIGO Scientific Collaboration and the Virgo Collaboration. LIGO Laboratory and Advanced LIGO are funded by the United States National Science Foundation (NSF) as well as the Science and Technology Facilities Council (STFC) of the United Kingdom, the Max-Planck-Society (MPS), and the State of Niedersachsen/Germany for support of the construction of Advanced LIGO and construction and operation of the GEO600 detector. Additional support for Advanced LIGO was provided by the Australian Research Council. Virgo is funded, through the European Gravitational Observatory (EGO), by the French Centre National de Recherche Scientifique (CNRS), the Italian Istituto Nazionale di Fisica Nucleare (INFN) and the Dutch Nikhef, with contributions by institutions from Belgium, Germany, Greece, Hungary, Ireland, Japan, Monaco, Poland, Portugal, Spain~\citep{Rich_Abbott_2021}. The authors are grateful for computational resources provided by the LIGO Laboratory and supported by National Science Foundation Grants PHY-0757058 and PHY-0823459.

\bibliography{biblio.bib}

\end{document}